\documentclass[aps,prb,reprint,superscriptaddress,english]{revtex4-2}
\usepackage[utf8]{inputenc}
\usepackage[T1]{fontenc}
\usepackage{amsmath,amssymb}
\usepackage{graphicx}
\usepackage{float}
\usepackage{mathptmx}
\usepackage{color}
\usepackage[colorlinks=true,citecolor=blue,linkcolor=blue,urlcolor=blue]{hyperref}

\definecolor{mygreen}{rgb}{0.0, 0.5, 0.0}

\begin{document}

\title{Rashba-controlled emergence and reversal of critical-field anisotropy in \(p\)-wave magnet--superconductor hybrids}

\author{Rui-Xiang Cheng}
\affiliation{School of Physical Science and Technology, Soochow University, Suzhou 215006, China}
	
\author{Zhao Gong}
\affiliation{School of Physical Science and Technology, Soochow University, Suzhou 215006, China}

\author{Dong-Hui Xu}
\affiliation{Department of Physics, Chongqing University, Chongqing 400044, China}
\affiliation{Chongqing Key Laboratory for Strongly Coupled Physics, Chongqing University, Chongqing 400044, China}
	
\author{Chui-Zhen Chen}
\email{czchen@suda.edu.cn}
\affiliation{School of Physical Science and Technology, Soochow University, Suzhou 215006, China}

\date{\today}

\begin{abstract}
We show that tuning Rashba spin-orbit coupling can change the in-plane field direction along which an $s$-wave superconductor is most susceptible to Zeeman depairing. We consider a two-dimensional $s$-wave superconductor subject to an odd-parity exchange field induced by a $p$-wave magnet. Without Rashba coupling, the out-of-plane exchange field enhances the in-plane critical field but leaves its angular response exactly isotropic, despite the anisotropic magnetic spin texture. Minimizing the Bogoliubov--de Gennes grand potential over uniform singlet pairing, we find that increasing Rashba coupling first generates and then reverses the principal-axis critical-field contrast, switching the minimum-critical-field direction by $90^{\circ}$. Equal principal-axis fields at the intervening compensation point coexist with residual angular modulation. We interpret the reversal through competing momentum-dependent depairing contributions from the magnetic and Rashba spin textures, while the on-site singlet pairing channel remains unchanged. Direction-dependent discontinuous collapses of the order parameter further shape the critical-field curves. These results identify a tunable directional paramagnetic response within the uniform-pairing sector. In heterostructures with sufficient electrostatic tunability of interfacial Rashba coupling, this reversal would enable gate control of the easy-depairing axis without reorienting the magnetic texture.
\end{abstract}

\maketitle

\section{INTRODUCTION}

The response of a superconductor to a magnetic field depends on both the structure of its Cooper pairs and the spin texture of the electronic states from which they form. Ising superconductors illustrate how spin locking can enhance magnetic resilience: an out-of-plane spin-orbit field suppresses in-plane Zeeman depairing and allows superconductivity to survive beyond the conventional Pauli limit~\cite{2016ising_NbSe2_np,Mockli2020_Ising_all_PRB}. This protection need not require a new pairing symmetry, although spin-dependent band structures and available pairing channels can strongly influence the superconducting response~\cite{Mockli2020_Ising_all_PRB,Sigrist1991_RMP}. A related question concerns direction rather than magnitude: can one switch the in-plane field direction that most efficiently destroys superconductivity while keeping the pairing interaction and the magnetic texture fixed?

This distinction is important when interpreting anisotropic superconducting responses. Electronic nematicity, charge ordering, and rotational-symmetry breaking have been investigated in several unconventional superconducting systems~\cite{chu2010_Iron_plane,Matano2016_SpinRotation_CuBiSe,Jiang2021_ChiralCDW_KV3Sb5,2014_nematic_order_np,2021cao_C2_science,2022charge_density_Kagome}. Strain, sample geometry, reduced dimensionality, and engineered interfaces also affect superconductivity and its critical fields~\cite{2021C2_NbSe2_strain_np,Hicks2014_Strain_Sr2RuO4,Moshchalkov1995_Topology_Superconductor,SaintJames1963_Onset_Superconductivity,Ruggiero1982_Superlattice_NbGe,Jiang1995_SuperconductingFerromagnetic_Superlattices}. An anisotropic field response therefore need not originate from an anisotropic pairing order parameter. In the setting considered here, the directional response is instead controlled by the competition between two normal-state spin-splitting mechanisms: magnetic exchange and Rashba spin-orbit coupling (SOC).

Compensated magnets provide a route to momentum-dependent spin splitting without a net magnetization. Altermagnets exhibit even-parity spin splitting~\cite{2022_Smejkal_AltermagnetReview_PRX}, with characteristic spectroscopic and transport signatures~\cite{2024_alter_ARPES_nature,2025_alter_AHE_PRB}. Their odd-parity counterparts, $p$-wave magnets, possess spin-resolved dispersions satisfying the effective time-reversal relation $E_{\mathbf{k}\uparrow}=E_{-\mathbf{k}\downarrow}$~\cite{2023_p_arxiv,2024_p_min_PRL,2025_p_eles_Nature,2026_pswitch_CeNiAsO}. Proposed responses include anisotropic resistivity~\cite{2023_p_arxiv}, tunneling magnetoresistance~\cite{2024_p_min_PRL}, and an out-of-plane spin polarization generated by the nonrelativistic Edelstein effect~\cite{2025_p_Edelstein_nc}. Magnetic superconducting hybrids provide a complementary setting in which the exchange texture acts on Cooper pairs~\cite{Mazanik2026_AltermagnetismSC}. In particular, an odd-parity exchange field directed out of plane can enhance the in-plane critical field of an $s$-wave superconductor with an on-site attractive interaction, a mechanism termed pseudo-Ising superconductivity~\cite{Sun2025_PseudoIsing}. This establishes a magnetic route to Ising-like protection~\cite{2026_Nonrelativistic_p_wave}; the question here is how its directional character can be reorganized.

An anisotropic exchange texture does not, by itself, imply an anisotropic in-plane critical field. For the purely out-of-plane exchange considered here, rotating the in-plane field is equivalent to a global spin rotation about the out-of-plane axis, and the Zeeman response is exactly isotropic in the absence of Rashba SOC. Interfacial Rashba coupling removes this equivalence by adding an in-plane spin field tied to momentum. The two contributions need not act as additive sources of protection: Rashba coupling can introduce a field-parallel component in an exchange-protected sector while supplying transverse spin locking near an exchange node. Whether these competing effects merely suppress the critical field or reverse its directional preference is therefore a nontrivial superconducting question.

Rashba-induced low-temperature order-parameter collapse has been studied in Ising superconductors~\cite{2025_pR_collapse}, and weakly spin-split Fermi-surface sectors can strongly influence their critical fields~\cite{Ising_field_symmetry}. Our focus is the reversal of directional depairing in a fixed anisotropic magnetic background, rather than the existence of spin-locking protection or Rashba-induced collapse alone. This reversal would allow the most field-sensitive direction of an on-site singlet condensate to be changed without rotating the magnetic texture or introducing an independent triplet order parameter.

Here we investigate a square-lattice model with $p$-wave exchange, Rashba SOC, and an on-site attractive interaction. Global minimization of the Bogoliubov--de Gennes (BdG) grand potential over uniform pairing amplitudes shows that increasing Rashba coupling reverses the principal-axis contrast from $H_{c2x}>H_{c2y}$ to $H_{c2x}<H_{c2y}$ and switches the minimum-critical-field direction by $\pi/2$. Equal principal-axis fields at compensation coexist with residual angular modulation. We also resolve direction-dependent discontinuous order-parameter collapses and identify the exchange-SOC parameter range supporting reversal. These are properties of the paramagnetic response within the uniform-pairing sector, not evidence of a spontaneously nematic pairing order. Electrostatic modulation of interfacial Rashba SOC~\cite{Caviglia2010_Rashba_Oxide} motivates a possible control mechanism, whose physical requirements and limitations are discussed near the end of the paper.

The remainder of this paper is organized as follows. Section II introduces the tight-binding model and the grand-potential minimization method. Section III presents the angular dependence and Rashba-driven reversal of the paramagnetic critical-field anisotropy, discusses the spin-texture interpretation and direction-dependent depairing, and maps the parameter range supporting the reversal. Section III E discusses the electrostatic controllability of interfacial Rashba SOC. Section IV summarizes the results.

\begin{figure}[htbp]
\centering
\includegraphics[width=\linewidth]{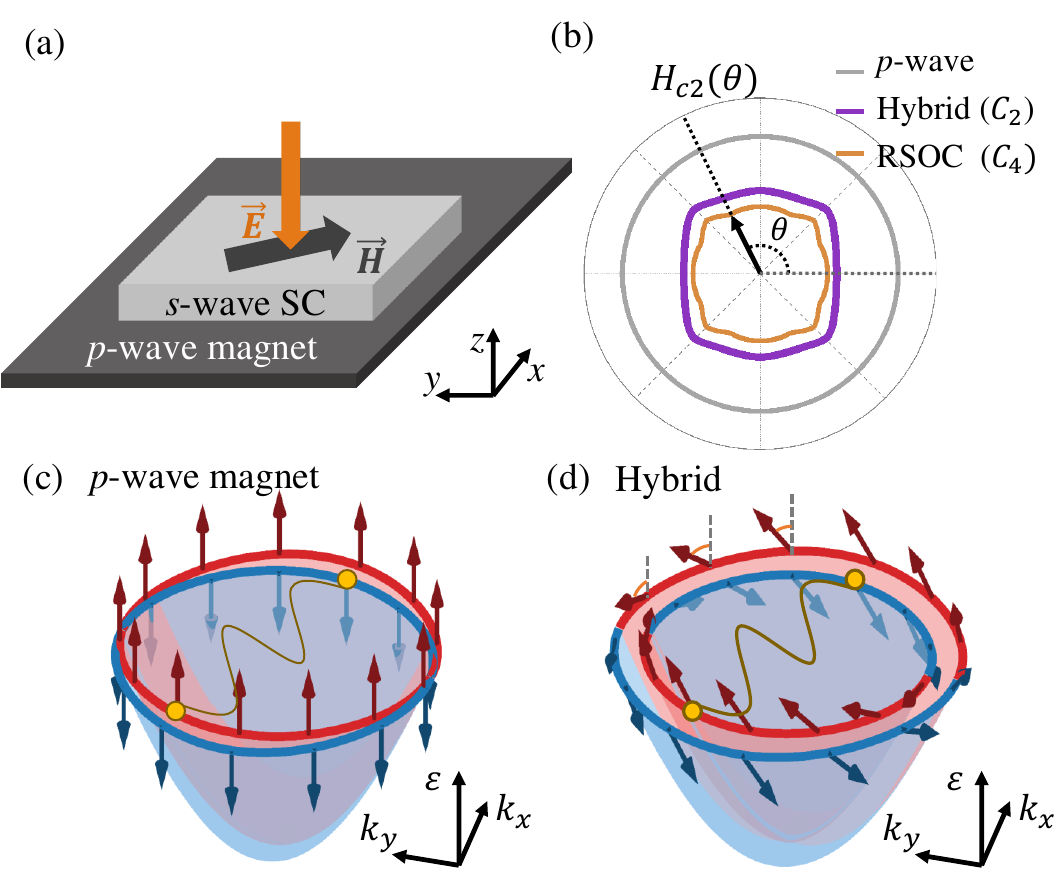}
\caption{Rashba-induced anisotropy in a \(p\)-wave-magnet and superconductor heterostructure.
\textbf{(a)} Conceptual device architecture: a 2D $s$-wave superconductor proximity-coupled to a $p$-wave magnet. The perpendicular electric-field arrow represents a possible gate-induced change of the confinement asymmetry and effective Rashba coupling, rather than an assumed universal relation $V_R \propto E_z$. An in-plane magnetic field \(\boldsymbol H\) probes the superconducting response; see Sec.~III~E. \textbf{(b)} Polar plot of the uniform-sector paramagnetic critical field $H_{c2}(\theta)$. The \(p\)-wave proximity effect alone yields an isotropic response (gray circle), whereas the Rashba-only case exhibits a \(C_4\) response in the model considered (orange curve). Their coexistence produces a \(C_2\) anisotropy (purple curve), reflecting the combined spin texture. \textbf{(c)} Momentum-space spin texture induced by the \(p\)-wave magnet. Out-of-plane spin locking provides pseudo-Ising protection against in-plane magnetic fields, yielding an isotropic \(H_{c2}\) despite the anisotropic spin texture.
\textbf{(d)} Hybridized spin texture in the presence of Rashba SOC. The Rashba field tilts the spins into the plane, and its competition with the \(p\)-wave exchange field generates a \(C_2\) angular dependence of \(H_{c2}\).}
\label{fig:device}
\end{figure}

\section{MODEL AND METHODS}
\label{sec:model}

\subsection{Model Hamiltonian and superconducting pairing}
To systematically investigate the interplay between odd-parity $p$-wave magnetism and Rashba spin-orbit coupling (SOC), we construct an effective tight-binding model on a two-dimensional square lattice. The normal-state Hamiltonian in momentum space is defined as:
\begin{equation}
\mathcal{H}_{N}(\mathbf{k})=h_{0}(\mathbf{k})+h_{J}(\mathbf{k})+h_{R}(\mathbf{k})+h_{Z}.
\end{equation}
The kinetic energy is given by $h_{0}(\mathbf{k})=-2t(\cos k_{x}+\cos k_{y})-\mu$, where $t$ is the nearest-neighbor hopping amplitude and $\mu$ is the chemical potential. The $p$-wave magnetic exchange field is modeled by the momentum-dependent term $h_{J}(\mathbf{k})=V_J\sin k_{x}\cos k_{y}\sigma_{z}$. Here, we emphasize that the $\cos k_{y}$ factor appearing in the effective $p$-wave magnetic Hamiltonian is a direct remnant originating from the itinerant nature and bond-dependent spatial dispersion of the underlying antiferromagnetic substrate~\cite{Sun2025_PseudoIsing}. A comprehensive discussion of this term is provided in Appendix~\ref{app:microscopic}. Crucially, while this parity-breaking term intrinsically possesses $C_{2}$ spatial anisotropy, it strictly preserves time-reversal symmetry (TRS) in momentum space, satisfying the standard relation $\mathcal{T} h_J(\mathbf{k}) \mathcal{T}^{-1} = h_J(-\mathbf{k})$ with $\mathcal{T}=i\sigma_y\mathcal{K}$. This effective relation should not be identified with unbroken microscopic time reversal of the parent magnetic order, whose protecting antiunitary symmetry may include a lattice translation~\cite{Sun2025_PseudoIsing,2026_Nonrelativistic_p_wave,Sigrist1991_RMP}. The Rashba SOC, $h_{R}(\mathbf{k})=V_R(\sin k_{x}\sigma_{y}-\sin k_{y}\sigma_{x})$, generates an in-plane helical spin texture. The competition between this Rashba field and the out-of-plane \(p\)-wave exchange texture controls the anisotropic superconducting response discussed below.

In the strictly two-dimensional limit under a parallel magnetic field, we neglect orbital depairing and retain only the Zeeman coupling
$h_{Z}=\mathbf{H}_{\parallel}\cdot\boldsymbol{\sigma}$, with $\mathbf{H}_{\parallel}=(H_{x},H_{y},0)$, where the Zeeman coefficient has been absorbed into \(\mathbf H_{\parallel}\), which therefore has energy units. For a scalar electronic $g$ factor, its magnitude is $H = g \mu_B B / 2$, where $B$ is the physical magnetic induction. We restrict our analysis to the uniform pairing sector (\(\mathbf q=0\)) with an on-site attractive interaction \(U_{\mathrm{sc}}>0\).
In the Nambu basis $\Psi_{\mathbf{k}}=(c_{\mathbf{k}\uparrow},c_{\mathbf{k}\downarrow},c_{-\mathbf{k}\uparrow}^{\dagger},c_{-\mathbf{k}\downarrow}^{\dagger})^{T}$, the $4\times4$ BdG Hamiltonian reads:
\begin{equation}
\mathcal{H}_{\text{BdG}}(\mathbf{k})=\begin{pmatrix}\mathcal{H}_{N}(\mathbf{k}) & -i\Delta\sigma_{y}\\
i\Delta\sigma_{y} & -\mathcal{H}_{N}^{*}(-\mathbf{k})
\end{pmatrix},
\end{equation}
where the superconducting order parameter is determined self-consistently as
$\Delta=\frac{U_{\mathrm{sc}}}{A}\sum_{\mathbf k}
\left\langle c_{-\mathbf k\downarrow}c_{\mathbf k\uparrow}\right\rangle $,
with \(A\) the number of lattice sites.
The lattice constant is set to unity.
Although the pairing interaction is purely on-site and spin singlet, the odd-in-momentum magnetic exchange and Rashba SOC induce odd-parity spin-triplet components in the anomalous correlations. The resulting state therefore consists of an \(s\)-wave singlet order parameter accompanied by induced \(p\)-wave triplet correlations, rather than an independent triplet order parameter~\cite{Sun2025_PseudoIsing,Sigrist1991_RMP,2026_Nonrelativistic_p_wave}.

\subsection{Grand-potential minimization and analytical spectra}
Near a first-order Pauli-limited transition, iterative solutions of the self-consistent gap equation can depend on the initial condition and converge to metastable states. Moreover, at low temperatures, the linearized gap equation determines only the local instability of the normal state which can differ from the thermodynamic transition field~\cite{1964_pauli_first_order}.
To capture the direction-dependent first-order Pauli collapse~\cite{Clogston1962_PauliLimit}, we therefore determine the equilibrium uniform state by globally minimizing the mean-field grand-potential density at fixed $\mu$. Up to a pairing-independent contribution, this is:
\begin{equation}
\label{eq:free_energy}
F = \frac{|\Delta|^2}{U_{sc}} - \frac{1}{2\beta A} \sum_{\mathbf{k},n} \ln\left(1 + e^{-\beta E_n(\mathbf{k})}\right),
\end{equation}
where $\beta = 1/(k_B T)$, $A$ is the number of sites, and \(n\) runs over all four BdG eigenvalues. The term omitted from Eq.~(\ref{eq:free_energy}) cancels in comparisons of $\delta F(\Delta)=F(\Delta)-F(0)$ at identical $\mu, T$, and field. We retain the symbol $F$ for this grand-potential density throughout. Choosing \(\Delta\) to be real, the stationary condition $\partial F / \partial \Delta = 0$ is equivalent to the conventional self-consistent gap equation.
Rather than relying on iterative convergence, we scan \(F(\Delta)\) and select its global minimum, thereby distinguishing the thermodynamically stable state from metastable extrema within the chosen variational sector. 

In our calculations, we set $t=1$, $\mu=-1$, and the pairing interaction $U_{sc} = 1.5$, yielding a zero-temperature gap $\Delta_0 \approx 0.07$ in the pristine state. All calculations of $F(\Delta)$ are performed on $1000\times1000$ $\mathbf{k}$-grid at a low temperature of $T = 0.03\,T_c$. Here $T_c$ denotes the zero-field critical temperature of the pristine reference system. We use $H_{c20}$ for its paramagnetic critical field at the same temperature; in the low-temperature weak-coupling limit, $H_{c20}=\Delta_0/\sqrt{2}$~\cite{Clogston1962_PauliLimit}. Throughout the paper, $H_{c2}(\theta)$ denotes the thermodynamic superconducting-to-normal transition field within the uniform ($\mathbf{q}=0$) pairing sector. It is neither an orbital upper critical field nor a critical field optimized over modulated superconducting states.

Before investigating the Rashba-induced critical-field anisotropy, we give the explicit analytical solution in the pure $p$-wave magnet limit, which provides the physical baseline for the full problem. 
The general solution with Rashba SOC is presented in
Appendix~\ref{app:analytical}. In the $V_R = 0$ limit, the positive quasiparticle excitation spectra analytically simplify to $E_{\pm}(\mathbf{k})$:
\begin{equation}
E_{\pm}(\mathbf{k}) = \sqrt{e_0^2 + e_J^2 + \Delta^2 + |\mathbf{H}|^2 \pm 2\sqrt{e_0^2 e_J^2 + |\mathbf{H}|^2 (e_0^2 + \Delta^2)}},
\end{equation}
where $e_0(\mathbf{k}) = -2t(\cos k_x + \cos k_y) - \mu$ and $e_J(\mathbf{k}) = V_J \sin k_x \cos k_y$. 
Crucially, because the external magnetic field enters the spectrum only through the rotationally invariant combination $|\mathbf{H}|^2$, both the quasiparticle spectrum and the uniform-state grand potential are independent of the in-plane field orientation. 
Thus, despite the \(C_2\)-symmetric momentum dependence of \(e_J(\mathbf{k})\), the pure \(p\)-wave magnetic state exhibits an exactly isotropic in-plane Zeeman response.

Finally, Rashba SOC and an in-plane field can favor finite-momentum helical pairing~\cite{Yip2014_NCS_Review,Edelstein1995_Magnetoelectric,Barzykin2002_InPlaneHc2,Kaur2005_HelicalPhase}. Determining the unrestricted equilibrium state therefore requires minimization over both $\Delta$ and $\mathbf{q}$. The present $\mathbf{q}=0$ calculation isolates the directional pair breaking generated by the competing magnetic and Rashba spin textures. The inequalities $V_J, V_R \ll t$ do not by themselves ensure that finite-momentum corrections are small relative to the principal-axis contrast. Our calculations establish the reversal within the uniform sector; whether its location, magnitude, or existence remains unchanged after optimization over $\mathbf{q}$ is not determined here.

\section{RESULTS AND DISCUSSION}
\label{sec:results}

\subsection{Angular dependence and anisotropy reversal}
\label{sec:angular}

\begin{figure}[htbp]
    \centering
    \includegraphics[width=0.47\textwidth]{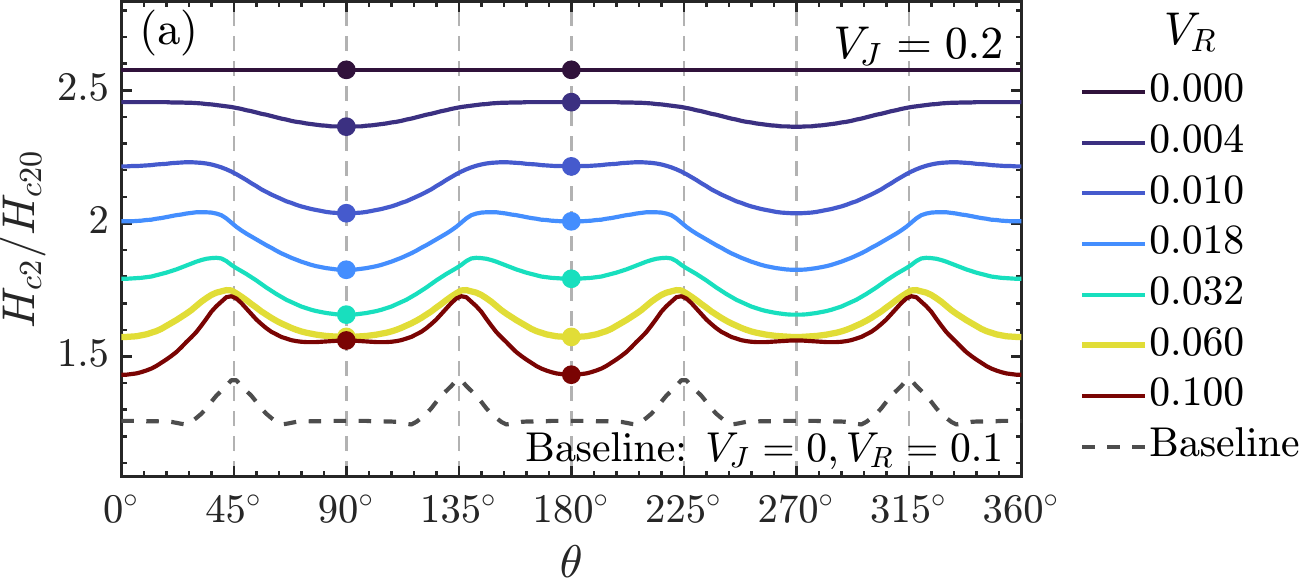}
    \\[0.3cm]
    \includegraphics[width=0.47\textwidth]{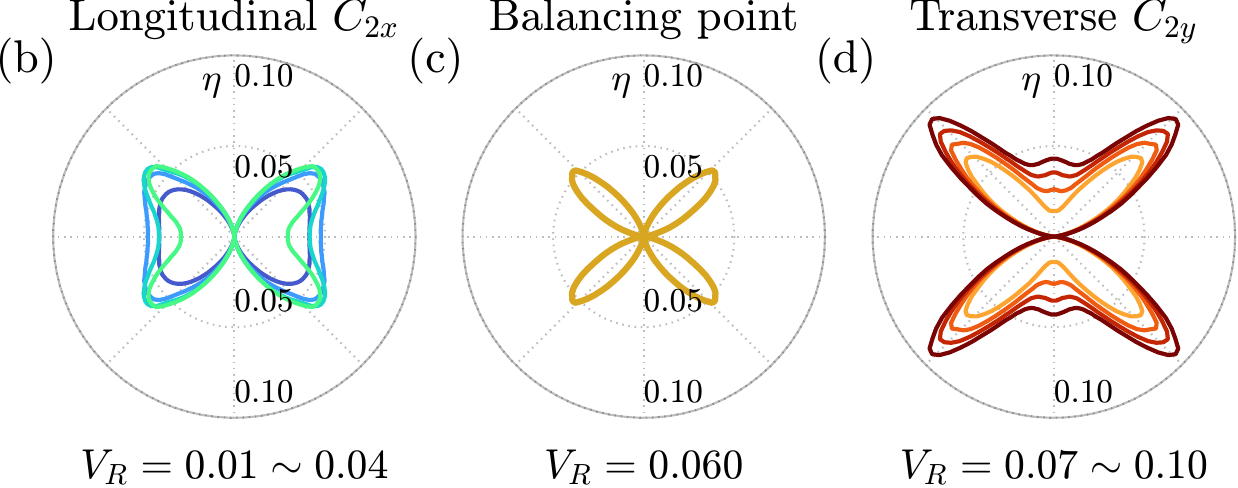}
\caption{Rashba-driven crossover from longitudinal to transverse critical-field anisotropy. \textbf{(a)} Angular dependence of the normalized in-plane critical field $H_{c2}/H_{c20}$ at fixed $p$-wave exchange $V_J = 0.2$ for different Rashba SOC strengths \(V_R\). Solid dots mark \(H_{c2x}\) and \(H_{c2y}\), highlighting the reversal of their relative magnitudes as \(V_R\) increases. The gray dashed curve shows the Rashba-only case ($V_J=0$, $V_R=0.1$). \textbf{(b)-(d)} Polar plots of the normalized angular variation $\eta(\theta)$ for three representative regimes. Increasing $V_R$ drives a continuous reversal of the principal-axis contrast from \textbf{(b)} a longitudinal \(C_{2x}\) anisotropy to \textbf{(d)} a transverse $C_{2y}$ anisotropy. This evolution passes through \textbf{(c)} an accidental anisotropy-compensation point ($V_R \approx 0.060$), where $H_{c2x} \simeq H_{c2y}$ without enforcing exact $C_4$ symmetry or angular isotropy. The $C_{2x}$ and $C_{2y}$ labels identify response orientations, not distinct pairing symmetries.}
\label{fig:angular}
\end{figure}

Figure~\ref{fig:angular} presents the angular dependence of the uniform-sector paramagnetic critical field $H_{c2}(\theta)$ as a function of the Rashba SOC strength $V_R$, with the $p$-wave magnetic exchange coupling fixed at $V_J = 0.2$. Here $\theta$ denotes the angle of the applied in-plane magnetic field relative to the principal $x$-axis. In the absence of Rashba SOC ($V_R=0$), the pure $p$-wave exchange significantly enhances the in-plane critical field to approximately 2.5 times the Pauli limit of the pristine $s$-wave reference system, while preserving exact in-plane isotropy. As \(V_R\) increases, Figure~\ref{fig:angular}(a) shows the emergence and subsequent reversal of the twofold anisotropy. To highlight its symmetry and orientation independently of the overall magnitude of \(H_{c2}\), Figs.~\ref{fig:angular}(b)-(d) display the normalized angular variation, 
\begin{equation}
    \eta(\theta) = \frac{H_{c2}(\theta) - H_{c2,\min}}{H_{c2}(\theta) + H_{c2,\min}},
\end{equation}
where \(H_{c2,\min}=\min_{\theta}H_{c2}(\theta)\) for each \(V_R\). The three polar plots represent characteristic below-compensation, compensation, and above-compensation regimes.

In the weak-Rashba regime [$V_R \sim 0.01\text{--}0.04$, Fig.~\ref{fig:angular}(b)], the response departs from the isotropic baseline and develops an \(x\)-elongated twofold anisotropy, with \(H_{c2x}>H_{c2y}\). We refer to this as the longitudinal \(C_{2x}\) anisotropy regime. This label describes the critical-field response rather than a superconducting phase or the full spin-space symmetry of the Hamiltonian. In this regime, the minimum-$H_{c2}$ (easy-depairing) axis lies along $y$, while the maxima can occur at off-axis angles.

With increasing \(V_R\), the principal-axis contrast \(\delta H_{c2}=H_{c2x}-H_{c2y}\) decreases continuously and becomes nearly zero at \(V_R\simeq0.060\), defining an accidental anisotropy-compensation point [Fig.~\ref{fig:angular}(c)]. This near equality of \(H_{c2x}\) and \(H_{c2y}\) results from the competition between the \(p\)-wave magnetic and Rashba spin textures. However, \(H_{c2x}\simeq H_{c2y}\) indicates only the cancellation of the principal-axis contrast: \(H_{c2}(\theta)\) remains angle dependent and exhibits off-axis maxima. Rotational isotropy is therefore not restored. Although the profile may appear approximately fourfold, the coexistence of exchange and Rashba coupling does not generally enforce $\pi/2$ periodicity of the response; $\pi$ periodicity is retained. Equal principal-axis fields also need not imply the vanishing of any one angular harmonic. The compensation point therefore marks a cancellation of the net \(x\)-\(y\) contrast rather than a complete disappearance of anisotropy.

Beyond the compensation point, \(\delta H_{c2}\) changes sign and the twofold anisotropy re-emerges with the opposite orientation. In the above-compensation regime [$V_R \sim 0.07\text{--}0.10$, Fig.~\ref{fig:angular}(d)], $H_{c2y} > H_{c2x}$, producing a \(y\)-elongated polar profile that we denote the transverse \(C_{2y}\) anisotropy regime. This continuous reversal of the principal-axis contrast is accompanied by a $90^\circ$ switch of the minimum-$H_{c2}$ axis. It lies along \(y\) in the weak-Rashba regime, becomes degenerate along \(x\) and \(y\) near compensation, and shifts to \(x\) above compensation. The three angular profiles therefore demonstrate a Rashba-controlled inversion of the easy-depairing axis within the uniform superconducting sector. 

\subsection{Anisotropy ratio and direction-dependent order-parameter collapse}
\label{sec:cutoff}

\begin{figure}[htbp]
\centering
\includegraphics[width=\linewidth]{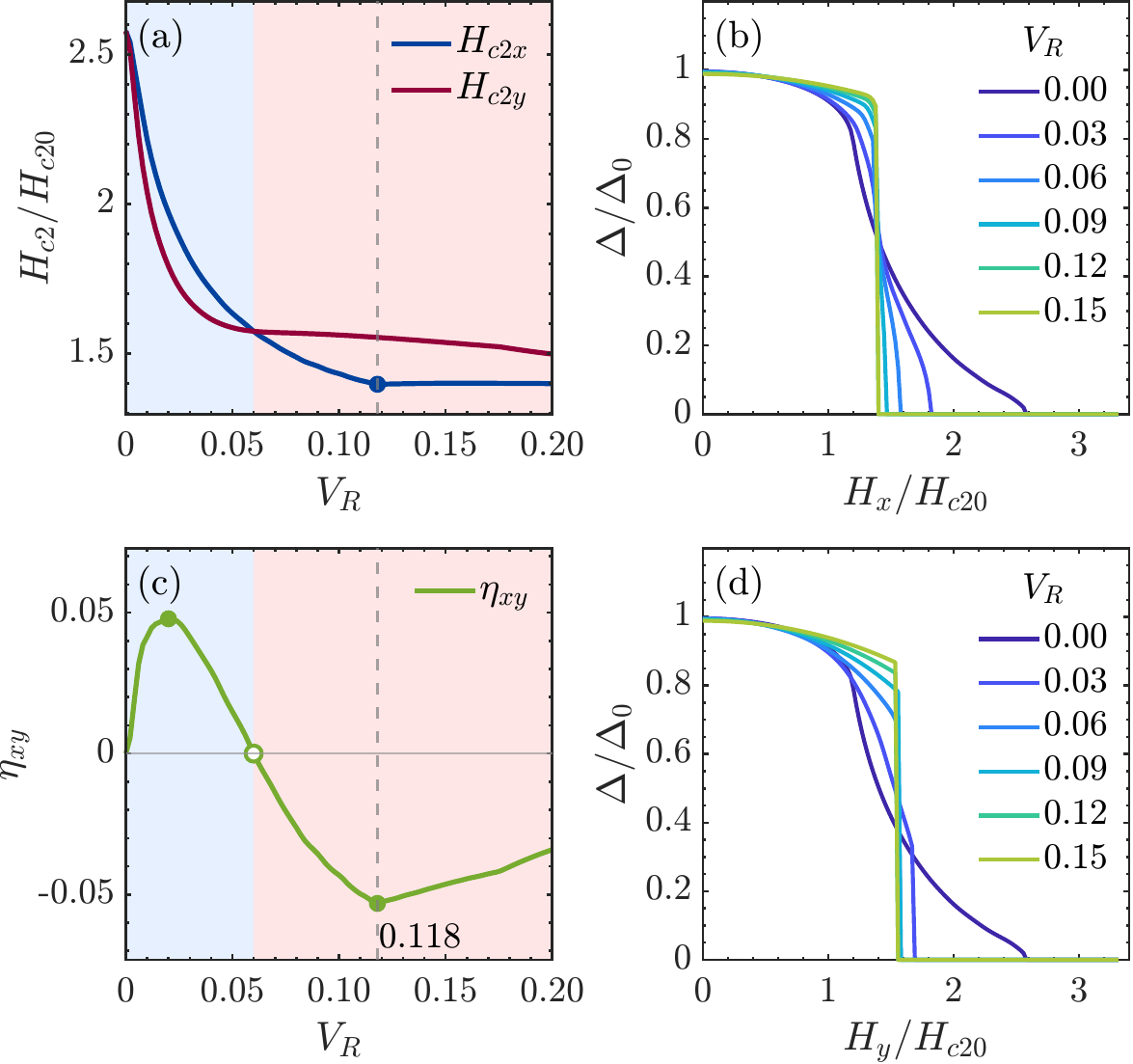}
\caption{Critical-field anisotropy and direction-dependent superconducting transitions. \textbf{(a)} Normalized critical fields $H_{c2x}$, $H_{c2y}$ and \textbf{(c)} the corresponding principal-axis contrast $\eta_{xy} = (H_{c2x}-H_{c2y})/(H_{c2x}+H_{c2y})$ versus Rashba SOC strength $V_R$ at $V_J = 0.2$. Background shading delineates the longitudinal $C_{2x}$ (blue) and transverse $C_{2y}$ (red) response regimes, separated by the accidental compensation point. A visible change in slope occurs near $V_R \approx 0.118$. \textbf{(b),(d)} Superconducting order parameter $\Delta$ versus normalized in-plane magnetic field along $x$ and $y$, respectively. The profiles reveal direction-dependent continuous suppression and discontinuous collapse of $\Delta$ switching discontinuously from a finite value to zero, accompanying the slope changes in (a). They do not directly plot the minimum quasiparticle excitation gap or establish a thermodynamic nonanalyticity in the coupling dependence.}
    \label{fig:gap}
\end{figure}

To quantitatively characterize the relative principal-axis critical fields, we introduce the dimensionless contrast parameter
\begin{equation}
    \eta_{xy} = \frac{H_{c2x} - H_{c2y}}{H_{c2x} + H_{c2y}},
\end{equation}
where \(H_{c2x}=H_{c2}(0)\) and \(H_{c2y}=H_{c2}(\pi/2)\). A positive \(\eta_{xy}\) means \(H_{c2x}>H_{c2y}\), whereas a negative value means \(H_{c2y}>H_{c2x}\). The location of the global angular minimum must additionally be established from $H_{c2}(\theta)$, as in Fig. 2; it does not follow from this two-axis ratio alone. The condition \(\eta_{xy}=0\) indicates compensation between the two principal-axis fields, but does not imply that the full angular dependence is isotropic. Figure~\ref{fig:gap} plots \(\eta_{xy}\), together with \(H_{c2x}\) and \(H_{c2y}\), as functions of \(V_R\).

In the weak-Rashba regime (light-blue shaded region), $\eta_{xy} > 0$, corresponding to \(H_{c2x}>H_{c2y}\) and the $y$-directed easy-depairing axis found in Fig.~\ref{fig:angular}; we refer to this as the longitudinal $C_{2x}$ regime. For fixed $V_J=0.2$, both principal-axis critical fields decrease overall with increasing $V_R$, reflecting the suppression of the uniform-state critical field by Rashba SOC. Meanwhile, \(\eta_{xy}\) evolves nonmonotonically: it first increases to a maximum and then decreases, changing sign near $V_R \approx 0.06$. This zero-crossing marks the accidental compensation identified in Fig.~\ref{fig:angular}. Beyond this point (light-red shaded region), $\eta_{xy}<0$, corresponding to \(H_{c2y}>H_{c2x}\) and the $x$-directed easy-depairing axis; this defines the transverse \(C_{2y}\) regime.

The otherwise smooth evolution of $\eta_{xy}$ shows a visible kink near $V_R \approx 0.12$, accompanied by a slope change of $H_{c2x}$ in Fig.~\ref{fig:gap}(a). To examine the associated depairing behavior, we consider the field-dependent order parameter $\Delta(H)$ in Figs.~\ref{fig:gap}(b) and (d). For $\mathbf H\parallel x$, suppression evolves from continuous vanishing at weak $V_R$ to an abrupt first-order collapse at larger $V_R$. The latter occurs when the superconducting and normal-state grand-potential minima become degenerate, with the equilibrium $\Delta$ switching discontinuously from a finite value to zero. The $\mathbf H\parallel y$ profiles develop an abrupt collapse at smaller $V_R$. This directional difference accompanies the slope changes in the critical-field curves. A change of transition order does not, by itself, establish a first-derivative nonanalyticity of $H_{c2}(V_R)$; the precise structure near the kink would require a more detailed resolution of the competing minima.

Throughout this discussion, $\Delta$ denotes the pairing amplitude, not the minimum quasiparticle excitation energy. The latter can vanish while $\Delta$ remains finite. We therefore distinguish order-parameter suppression or collapse from quasiparticle-gap closing; an example follows directly from Eq. (4) at an exchange node, as discussed in the Appendix.

\subsection{Spin-texture competition and pairing protection}
\label{sec:mechanism}

\begin{figure}[htbp]
\centering
\includegraphics[width=\linewidth]{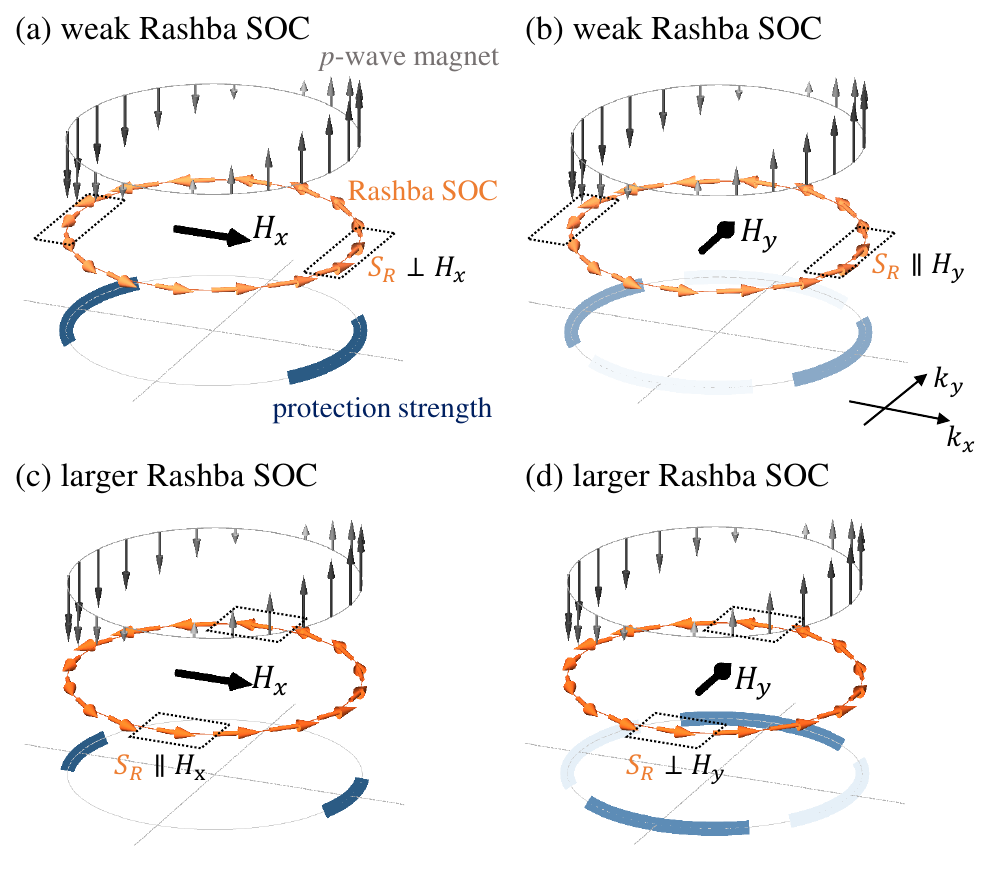}
\caption{Schematic illustration of Rashba-induced redistribution of $p$-wave magnetic protection. The three layers display the $p$-wave spin-splitting field (top), the Rashba field $S_R$ and applied in-plane Zeeman field $\mathbf{H}$ (middle), and the qualitative distribution of protection against Zeeman depairing (bottom). \textbf{(a), (b)} Weak and \textbf{(c), (d)} larger Rashba SOC, with $\mathbf{H}\parallel\hat{x}$ in (a), (c) and $\mathbf{H}\parallel\hat{y}$ in (b), (d). Dashed boxes highlight selected regions near the \(k_x\) axis, where the exchange provides spin-locking protection, and near the \(k_y\) axis, where the exchange vanishes. Rashba components parallel to $\mathbf{H}$ weaken the transverse character of the combined spin locking, whereas perpendicular components can reinforce protection or supply it near exchange nodes. The sketch is a qualitative illustration, not a calculated momentum-resolved pairing kernel; it omits other exchange nodes of $\sin k_x \cos k_y$. These contrasting local effects are consistent with the calculated reversal from $H_{c2x}>H_{c2y}$ to $H_{c2y}>H_{c2x}$.}
\label{fig:mechanism}
\end{figure}

The reversal to \(H_{c2y}>H_{c2x}\), together with the associated directional difference in Pauli depairing, admits a qualitative interpretation in terms of the momentum-dependent competition between the $p$-wave spin-splitting field $\mathbf{S}_{J}(\mathbf{k})$ and Rashba effective field $\mathbf{S}_{R}(\mathbf{k})$. We define $ \mathbf{S}_{J}=(0,0,V_J\sin k_x\cos k_y)$ and $\mathbf{S}_{R} = V_R(-\sin k_y,\sin k_x,0)$. As illustrated schematically in Fig.~\ref{fig:mechanism}, an internal spin-locking field transverse to the applied Zeeman field can reduce the pair-breaking effect of the field. By contrast, a parallel component generates an opposite-momentum energy mismatch and can weaken this protection.

In the absence of Rashba SOC, $\mathbf{S}_J$ points entirely out of plane and is perpendicular to any in-plane field. This transverse spin locking accounts for the enhanced critical field relative to the pristine $s$-wave state. Because the protection is equivalent for $\mathbf{H}_x$ and $\mathbf{H}_y$, the pure $p$-wave magnetic state remains macroscopically isotropic. Its strength is strongly momentum dependent: $\mathbf{S}_J$ is appreciable near parts of the $k_x$ axis but vanishes along $k_x=0$, producing inequivalent exchange-protected and exchange-node sectors.

In the weak-Rashba regime, one illustrative contribution comes from the exchange-protected sector near $k_y=0$ [dashed boxes in Figs.~\ref{fig:mechanism}(a) and (b)], where $\mathbf{S}_R \simeq (0,V_R\sin k_x,0)$. For an applied $\mathbf{H}_x$, both $\mathbf{S}_J$ and $\mathbf{S}_R$ are transverse to the field, preserving the original $p$-wave protection and supplementing it with transverse Rashba locking. For $\mathbf{H}_y$, however, the Rashba field is parallel to the applied field, tilting the combined spin-locking field toward it and weakening protection. This local contrast favors the $x$-field response, consistent with the calculated inequality $H_{c2x}>H_{c2y}$.

At larger $V_R$, a competing contribution is apparent near the $k_y$ axis [dashed boxes in Figs.~\ref{fig:mechanism}(c) and (d)], where the exchange is weak. Along $k_x=0$, $\mathbf{S}_J=0$ and $\mathbf{S}_R=(-V_R\sin k_y,0,0)$. For $\mathbf{H}_x$, the Rashba field is parallel to the applied field and supplies no transverse spin-locking protection, leaving this exchange-node sector susceptible to Zeeman depairing. For $\mathbf{H}_y$, the perpendicular Rashba field instead provides protection where the magnetic exchange alone offers none. Increasing $V_R$ strengthens this transverse locking and expands the surrounding region in which Rashba SOC competes with exchange. This sector therefore contributes a directional preference opposite to that of the exchange-protected sector.

The selected sectors in Fig.~\ref{fig:mechanism} are not exhaustive. The exchange also vanishes on $\cos k_y = 0$, and the relative contributions depend on the actual spin-split Fermi surfaces, pairing overlaps, and the full BdG spectrum. The grand-potential calculation includes all momenta, but the schematic does not provide a quantitative decomposition of the reversal. In a pure-Rashba system on the square lattice, the two principal axes are symmetry related; the exchange makes the competing sectors inequivalent.

\subsection{Parameter dependence of anisotropy reversal}
\label{sec:phasediagram}

\begin{figure}[htbp]
\centering
\includegraphics[width=\linewidth]{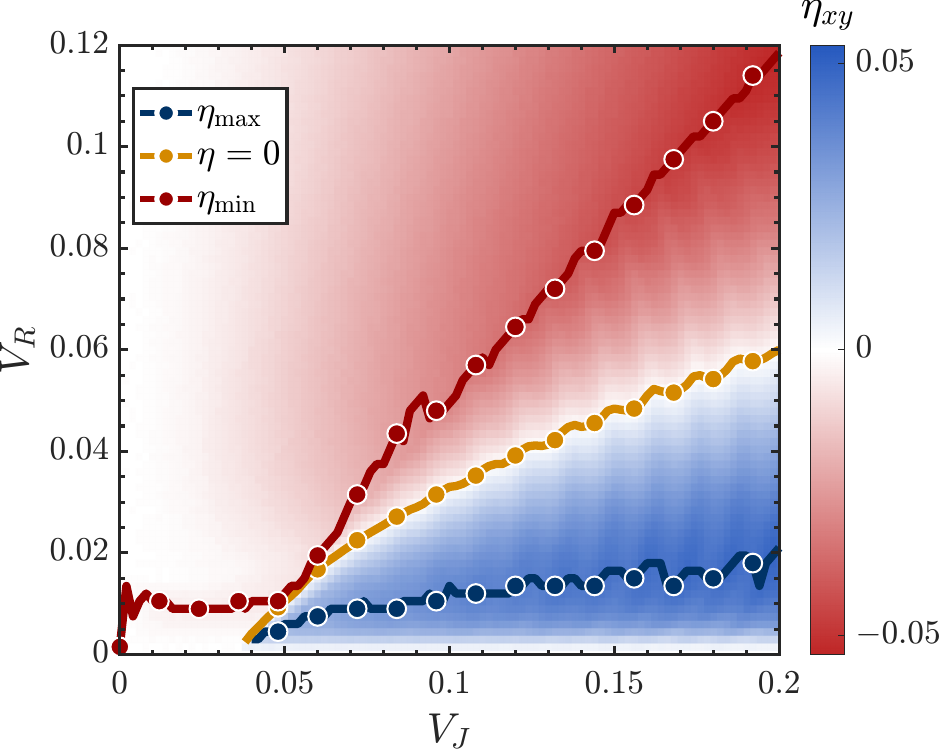}
\caption{Critical-field anisotropy in the $V_J$--$V_R$ plane for the uniform superconducting state. The color map shows $\eta_{xy}=(H_{c2x}-H_{c2y})/(H_{c2x}+H_{c2y})$. The golden curve  marks the compensation contour $\eta_{xy}=0$, separating the blue region with $H_{c2x}>H_{c2y}$ from the red region with $H_{c2y}>H_{c2x}$. The dark-blue  and the dark-red curves trace the maximum ($\eta_{\mathrm{max}}$) and minimum ($\eta_{\mathrm{min}}$) values of $\eta_{xy}$ as $V_R$ varies at fixed $V_J$. All three curves use circular markers. Equal principal-axis fields on the compensation contour can coexist with residual angular anisotropy. This is a response map, not a phase diagram of distinct pairing symmetries.}
\label{fig:phasediagram}
\end{figure}

The reversal of the in-plane critical-field anisotropy occurs over a range of exchange strengths, with the positive-contrast region becoming resolved near $V_J\simeq0.04$ for the parameters studied. Figure~\ref{fig:phasediagram} summarizes this behavior through $\eta_{xy}$ in the $V_J$--$V_R$ plane, calculated for the uniform superconducting state.

The compensation contour $\eta_{xy}=0$ separates the regimes with $H_{c2x}>H_{c2y}$ and $H_{c2y}>H_{c2x}$. Weaker Rashba SOC or stronger exchange favors the former, whereas stronger Rashba SOC or weaker exchange favors the latter. At each $V_J$, the $\eta_{\mathrm{max}}$ and $\eta_{\mathrm{min}}$ trajectories locate the largest positive and negative contrasts as $V_R$ varies. The compensation contour marks equality of the two principal-axis fields, while residual anisotropy may remain in the full angular profile. It therefore characterizes a reversal of the response anisotropy rather than a thermodynamic phase transition.

For weaker but finite exchange, $0<V_J\le0.04$, no region with $H_{c2x}>H_{c2y}$ is resolved within the sampled finite-$V_R$ range. This correlates with the exchange dependence of the pure $p$-wave critical field: its enhancement is marginal at weak exchange but becomes approximately linear at larger $V_J$. Thus, the regime with $H_{c2x}>H_{c2y}$ appears where exchange already produces an appreciable critical-field enhancement. This is consistent with weak Rashba SOC modifying the existing protection differently for the two directions, suppressing $H_{c2y}$ more strongly than $H_{c2x}$. The parameter map establishes where this initial anisotropy subsequently reverses to $H_{c2y}>H_{c2x}$ as $V_R$ increases. The sampled crossover near $V_J \simeq 0.04$ should not be interpreted as an exact threshold for the mechanism.

\subsection{Electrostatic control of interfacial Rashba coupling}
\label{sec:electrostatic}

The interfacial origin of Rashba SOC and its electrostatic tunability are distinct but compatible. The interface provides a built-in inversion-asymmetric environment. A gate can modify the confinement potential and the spatial and orbital character of the conducting states, thereby changing their effective Rashba coupling rather than the atomic SOC itself. Electric-field modulation of Rashba SOC has been demonstrated at the superconducting $\mathrm{LaAlO}_3/\mathrm{SrTiO}_3$ interface~\cite{Caviglia2010_Rashba_Oxide}. Microscopic studies also show that interfacial spin splitting can depend strongly on orbital hybridization~\cite{AlO3}. Accordingly, the electric-field arrow in Fig.~\ref{fig:device}(a) represents a possible control mechanism, not a material-independent proportionality between the applied field and $V_R$.

A useful phenomenological description is $V_R = V_R^{(0)} + \delta V_R(D_\perp, n)$ where $V_R^{(0)}$ is the built-in interfacial contribution, $D_\perp$ parametrizes the perpendicular electrostatic asymmetry, and $n$ is the carrier density. A candidate architecture is an electrostatically accessible ultrathin singlet-superconducting layer in contact with a fixed-domain $p$-wave magnet and controlled through a dielectric-separated gate. Its viability depends on screening and the gate response of the active states. In particular, a thick conducting layer placed between a gate and the superconducting channel can screen the intended modulation. The accessible $V_R$ range must therefore be determined for the specific heterostructure.

Separating changes in confinement asymmetry from carrier doping would help identify the role of Rashba SOC. Dual-gate control of spin-orbit interaction at approximately fixed carrier density has been demonstrated in InAs nanosheets~\cite{electrically_Rashba}, providing a control principle rather than a realization of the present magnetic superconducting hybrid. Where both gates have electrostatic access to the active layer, compensating gate voltages can vary the perpendicular asymmetry while reducing density changes. Constant-density control is nevertheless not equivalent to the fixed-$\mu$ sweep used here. A quantitative treatment of such a trajectory requires determining $\mu$ from the density constraint and accounting for any gate dependence of magnetic proximity and superconducting parameters. These device-specific calculations are not part of the present model.

For the angular responses found here, electrical reversal of the easy-depairing axis requires a gate trajectory to cross the compensation contour separating $H_{c2x} > H_{c2y}$ from $H_{c2x} < H_{c2y}$, with the full angular profile confirming the position of the minimum. The magnetic texture need not rotate: the gate changes the relative contributions of out-of-plane exchange and in-plane Rashba locking. At $V_J = 0.2t$, compensation occurs near $V_R \simeq 0.060t$ for the parameters of Fig.~\ref{fig:angular}. This is a model-dependent coupling scale, not a gate-voltage prediction or universal ratio. An interface whose built-in coupling is already near compensation would require only a smaller modulation across it, although a measurable directional contrast requires operating sufficiently far on either side.

An additional constraint follows directly from the Hamiltonian: at fixed remaining parameters, $H_{c2}(\theta; V_R) = H_{c2}(\theta; -V_R)$. A spin rotation by $\pi$ about $z$ maps $(V_R, \mathbf{H}_\parallel)$ to $(-V_R, -\mathbf{H}_\parallel)$ while leaving the exchange and singlet pairing invariant. The effective time-reversal relation then ensures equality of the grand potentials under reversal of the applied field. Thus, reversing the Rashba sign alone at unchanged magnitude does not switch the axis in this model. The proposed control relies on changing its magnitude relative to the other relevant scales and crossing the compensation contour. The perpendicular gate field itself need not select an in-plane direction; the fixed exchange texture supplies that directional reference.

The most direct observable is the gate evolution of the principal-axis contrast and the angle-resolved critical-field minimum, with the magnetic domain orientation held fixed. A change in the absolute critical field alone is less specific because gating can modify other parameters. Likewise, equal effective spin-splitting textures can give the same BdG response irrespective of their microscopic origin, so this observable is not a unique magnetic fingerprint. The present results establish a Rashba-controlled reversal within the uniform-pairing sector and identify conditions for an electrostatic realization. A quantitative device prediction additionally requires a calibrated gate-to-coupling relation and an assessment of finite-momentum pairing.

\section{CONCLUSION}
\label{sec:conclusion}
We have shown that the competition between odd-parity magnetic exchange and Rashba SOC can reverse the directional paramagnetic response of a two-dimensional singlet superconductor. Within the uniform-pairing sector, out-of-plane exchange enhances the in-plane critical field while preserving exact angular isotropy. Rashba coupling first generates a directional contrast and then reverses the ordering of the two principal-axis fields, switching the easy-depairing axis without changing the on-site singlet pairing channel or rotating the magnetic texture. 

Equal principal-axis fields at compensation coexist with residual angular modulation. Grand-potential minimization also resolves direction-dependent discontinuous order-parameter collapses, while the exchange-SOC map identifies the parameter range supporting reversal. The result is a controllable anisotropic superconducting response rather than a new pairing symmetry.

Electrostatic modulation of interfacial Rashba coupling provides a possible realization when the accessible gate trajectory crosses compensation. Its quantitative implementation and the equilibrium response beyond uniform pairing remain to be established.

\section*{Acknowledgements}
We acknowledge helpful discussions with Qi-Sheng Xu.
This project is supported by the National Key R\&D Program of China 
(Grant No.~2022YFA1403700), the National Natural Science Foundation 
of China (Grant Nos.~12074108 and 12347101), the Chongqing Natural 
Science Foundation (Grant No.~CSTB2022NSCQ-MSX0568), the Fundamental 
Research Funds for the Central Universities (Grant No.~2023CD-JXY048).

% ==============================================================================
% Appendices Section
% ==============================================================================
%\newpage

\appendix
\onecolumngrid % <--- 就是这个魔法指令，让接下来的内容全部变成单栏

\setcounter{figure}{0}
\renewcommand{\thefigure}{S\arabic{figure}}
\setcounter{equation}{0}
\renewcommand{\theequation}{S\arabic{equation}}
% -------------------------------------------

\section{Analytical Diagonalization and Landau Grand-Potential Landscapes}
\label{app:analytical}
\subsection{Normal-State Formalism and the BdG Matrix}
The normal-state Hamiltonian $\mathcal{H}_{N}(\mathbf{k})$ on a two-dimensional square lattice includes the kinetic energy, the $p$-wave magnetic exchange field, Rashba spin-orbit coupling (SOC), and an in-plane Zeeman field:
\begin{equation}
\mathcal{H}_{N}(\mathbf{k}) = h_{0}(\mathbf{k}) + h_{J}(\mathbf{k}) + h_{R}(\mathbf{k}) + h_{Z}.
\end{equation}
The kinetic term is $h_{0}(\mathbf{k}) = -2t(\cos k_{x} + \cos k_{y}) - \mu$, where $t$ is the nearest-neighbor hopping amplitude and $\mu$ is the chemical potential. The $p$-wave magnetic exchange term is $h_{J}(\mathbf{k}) = V_{J}\gamma(\mathbf{k})\sigma_{z}$, with the form factor $\gamma(\mathbf{k}) = \sin k_{x}\cos k_{y}$. This odd-in-momentum exchange term preserves time-reversal symmetry (TRS) in the effective model while breaking spatial inversion symmetry. The Rashba SOC is $h_{R}(\mathbf{k}) = V_{R}(\sin k_{x}\sigma_{y} - \sin k_{y}\sigma_{x})$, and the in-plane Zeeman term is $h_{Z} = \mathbf{H}\cdot\boldsymbol{\sigma}$, with $\mathbf{H} = (H_{x}, H_{y}, 0)$.

In the Nambu basis $\Psi_{\mathbf{k}} = (c_{\mathbf{k}\uparrow}, c_{\mathbf{k}\downarrow}, c_{-\mathbf{k}\uparrow}^{\dagger}, c_{-\mathbf{k}\downarrow}^{\dagger})^{T}$, we define the spin-dependent diagonal energies $e_{1,2} = h_{0}(\mathbf{k}) \pm V_{J}\gamma(\mathbf{k})$, the Zeeman amplitude $\beta_{1} = H_{x} - iH_{y}$, and the Rashba amplitude $\beta_{2} = V_{R}(i\sin k_{x} - \sin k_{y})$. Choosing the superconducting order parameter $\Delta$ to be real, the $4\times 4$ Bogoliubov--de Gennes (BdG) Hamiltonian reads:
\begin{equation}
\mathcal{H}_{\text{BdG}}(\mathbf{k}) = \begin{pmatrix}
e_{1} & \beta_{1} + \beta_{2}^{*} & 0 & -\Delta \\
\beta_{1}^{*} + \beta_{2} & e_{2} & \Delta & 0 \\
0 & \Delta & -e_{2} & -\beta_{1}^{*} + \beta_{2} \\
-\Delta & 0 & -\beta_{1} + \beta_{2}^{*} & -e_{1}
\end{pmatrix}.
\end{equation}

\subsection{Exact Solution and Macroscopic Isotropy in the Pure $p$-wave Limit ($V_R = 0$)}

Setting $V_R = 0$ recovers the pure $p$-wave magnetic limit, which provides an analytical baseline for the isotropic in-plane superconducting response.
Without Rashba SOC, the characteristic equation $\det(\mathcal{H}_{\text{BdG}} - E\mathbb{I}) = 0$ reduces to:
\begin{equation}
(E^2 - e_2^2 - \Delta^2)(E^2 - e_1^2 - \Delta^2) - 2|\mathbf{H}|^2(E^2 + e_1 e_2) + |\mathbf{H}|^4 - 2|\mathbf{H}|^2\Delta^2 = 0.
\end{equation}
Defining $L = E^2$, $e_0 = h_0(\mathbf{k})$, and $e_J = V_J\gamma(\mathbf{k})$, and using $e_1 e_2 = e_0^2 - e_J^2$ and $e_1^2 + e_2^2 = 2e_0^2 + 2e_J^2$, we obtain:
\begin{equation}
L_{\pm} = e_0^2 + e_J^2 + \Delta^2 + |\mathbf{H}|^2 \pm 2\sqrt{e_0^2 e_J^2 + |\mathbf{H}|^2 (e_0^2 + \Delta^2)}.
\end{equation}
The four BdG eigenvalues are therefore:
\begin{equation}
E_{1,2}(\mathbf{k}) = \pm \sqrt{L_{+}}, \quad E_{3,4}(\mathbf{k}) = \pm \sqrt{L_{-}}.
\end{equation}
Because the magnetic field enters only through $|\mathbf{H}|^2$, both the quasiparticle spectrum and the uniform-state grand potential are independent of the in-plane field orientation. This establishes an exactly isotropic in-plane Zeeman response in the absence of Rashba SOC.

For $\Delta\neq0$, the stationarity condition $\partial F/\partial\Delta=0$ yields the self-consistent gap equation:
\begin{equation} \label{eq:full_gap}
\frac{4A}{U_{sc}} = \sum_{\mathbf{k}, \alpha=\pm} \frac{\tanh(\beta E_{\alpha}/2)}{E_{\alpha}} \left( 1 \pm \frac{|\mathbf{H}|^2}{\sqrt{e_0^2 e_J^2 + |\mathbf{H}|^2 (e_0^2 + \Delta^2)}} \right),
\end{equation}
where the signs correspond to the branches $E_{\pm}=\sqrt{L_{\pm}}$. In our numerical calculations, we determine the equilibrium pairing amplitude by globally minimizing the grand-potential functional $F(\Delta)$ [Eq.~(\ref{eq:free_energy})], thereby distinguishing the thermodynamically stable state from metastable stationary solutions. This analytical formulation recovers the standard BCS, Pauli paramagnetic, and zero-field $p$-wave limits.

\subsection{General Solution with Rashba SOC via Cardano's Method ($V_R \neq 0$)}

When Rashba SOC is introduced, its interplay with the $p$-wave magnetic exchange produces an anisotropic in-plane superconducting response. The characteristic equation $\det(\mathcal{H}_{\text{BdG}} - E\mathbb{I}) = 0$ yields the quartic equation:
\begin{equation}
E^{4} + \alpha E^{2} + \beta E + \gamma = 0,
\end{equation}
with coefficients:
\begin{align}
\alpha &= -2(e_{0}^{2} + e_{J}^{2} + |\mathbf{H}|^{2} + V_{R}^{2} s_{\mathbf{k}}^{2} + \Delta^{2}), \\
\beta &= -8e_{0}D, \\
\gamma &= (e_{0}^{2} - e_{J}^{2} - |\mathbf{H}|^{2} - V_{R}^{2} s_{\mathbf{k}}^{2})^{2} - 4D^{2} \nonumber \\
&\quad + 2\Delta^{2}(e_{0}^{2} + e_{J}^{2} - |\mathbf{H}|^{2} + V_{R}^{2} s_{\mathbf{k}}^{2}) + \Delta^{4},
\end{align}
where we have defined $s_{\mathbf{k}}^2 \equiv \sin^2 k_x + \sin^2 k_y$ and $D = V_{R}(H_{y} \sin k_{x} - H_{x} \sin k_{y})$.

The quartic is solved via Cardano's method. The resolvent cubic is:
\begin{equation}
P^{3} + 2\alpha P^{2} + (\alpha^{2} - 4\gamma)P - \beta^{2} = 0.
\end{equation}
With $P = P' - 2\alpha/3$, we obtain the depressed cubic $P'^{3} + A_{1}P' + A_{2} = 0$, where $A_{1} = -\alpha^{2}/3 - 4\gamma$ and $A_{2} = -2\alpha^{3}/27 + 8\gamma\alpha/3 - \beta^{2}$. For our parameter regime ($t=1$, $\mu=-1$, $V_J \le 0.4$, $V_R \le 0.3$), we have numerically verified that $\alpha^2+12\gamma>0$ throughout the Brillouin zone. We choose the largest nonnegative real root of the resolvent cubic, given by:
\begin{equation}
P = \frac{2}{3} \left( \sqrt{\alpha^{2} + 12\gamma} \cos \Theta - \alpha \right),
\end{equation}
with $\Theta = \frac{1}{3} \arccos \left[ (2\alpha^{3} - 72\gamma\alpha + 27\beta^{2}) / 2(\alpha^{2} + 12\gamma)^{3/2} \right]$. The four BdG eigenvalues are then expressed in two pairs, with the correlated signs written explicitly:
\begin{align}
E_{1,2}(\mathbf{k}) &= \frac{1}{2} \left( \sqrt{P} \pm \sqrt{-\left(2\alpha + P + \frac{2\beta}{\sqrt{P}}\right)} \right), \label{eq:E12} \\
E_{3,4}(\mathbf{k}) &= \frac{1}{2} \left( -\sqrt{P} \pm \sqrt{-\left(2\alpha + P - \frac{2\beta}{\sqrt{P}}\right)} \right). \label{eq:E34}
\end{align}

Equations~(\ref{eq:E12}) and (\ref{eq:E34}) apply when $P > 0$. At a degenerate $P = 0$ limit, the quartic is biquadratic ($\beta = 0$) and is solved directly as $E^2 = [-\alpha \pm \sqrt{\alpha^2 - 4\gamma}]/2$, avoiding division by zero. This direct biquadratic solution is also available whenever $\beta = 0$. Degenerate limits of the arccosine representation can likewise be handled directly through the polynomial.

With the analytical eigenvalues $E_n(\mathbf{k})$ determined, we evaluate the thermodynamic grand-potential functional [Eq.~(\ref{eq:free_energy})]. Iterative self-consistent gap solvers can converge to metastable solutions, particularly near first-order transitions. We therefore determine the equilibrium pairing amplitude within the uniform pairing sector by scanning the grand-potential landscape and locating its global minima, as described in the main text and illustrated in Fig.~\ref{fig:free_energy_profile}.

\subsection{First-Order Transitions and Grand-Potential Profiles}

\begin{figure}[htbp]
\centering
\includegraphics[width=0.8\linewidth]{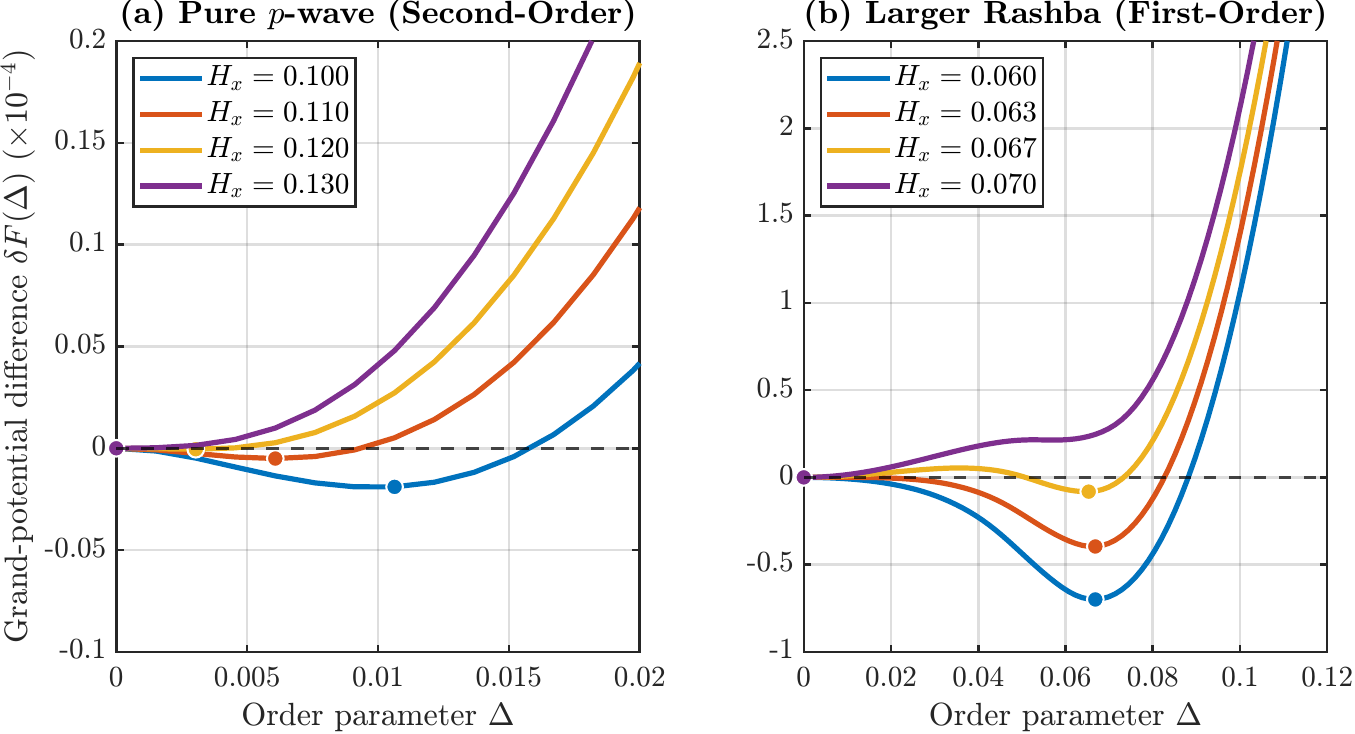}
\caption{Evolution of the grand-potential profile $\delta F(\Delta)$ under varying in-plane magnetic fields.}
\label{fig:free_energy_profile}
\end{figure}

Using the exact spectra $E_n(\mathbf{k})$, we determine the upper critical field $H_{c2}$ and the transition order from the grand-potential landscape. Figure~\ref{fig:free_energy_profile} shows the grand-potential difference $\delta F(\Delta)=F(\Delta)-F(0)$ at different in-plane magnetic fields, illustrating the continuous vanishing of the equilibrium pairing amplitude at a second-order transition and its discontinuous collapse from a finite value to zero at a first-order Pauli transition.

A discontinuity of the minimizing $\Delta$ establishes a first-order transition within this variational space. A visually sharp variation of $\Delta(H)$ alone is not the thermodynamic criterion; the superconducting and normal minima must be compared. Likewise, the change of transition order does not alone establish a nonanalytic first derivative of $H_{c2}(V_R)$. For a regular local expansion $\delta F=a\Delta^2+b\Delta^4+c\Delta^6$ with $b<0$ and $c>0$, first-order coexistence occurs at $a=b^2/(4c)$ and $\Delta_{\mathrm{jump}}^2=-b/(2c)$. This illustrates why the precise coupling dependence near the apparent kink cannot be inferred solely from the existence of discontinuous order-parameter collapse.

The order parameter is also distinct from the minimum quasiparticle excitation gap. At an exchange node and $V_R=0$, Eq.~(S4) gives $E_{-}(\mathbf{k}) = |\sqrt{e_0(\mathbf{k})^2+\Delta^2} - H|$. Thus, quasiparticle gaplessness can occur while $\Delta$ remains finite. All $\Delta$ profiles and maps below refer to the pairing amplitude, not directly to the minimum excitation energy.

\section{Global $\Delta$-$T$-$H$ Thermodynamic Phase Diagrams}
\label{app:phasediagrams}

\begin{figure}[htbp]
\centering
\includegraphics[width=0.95\linewidth]{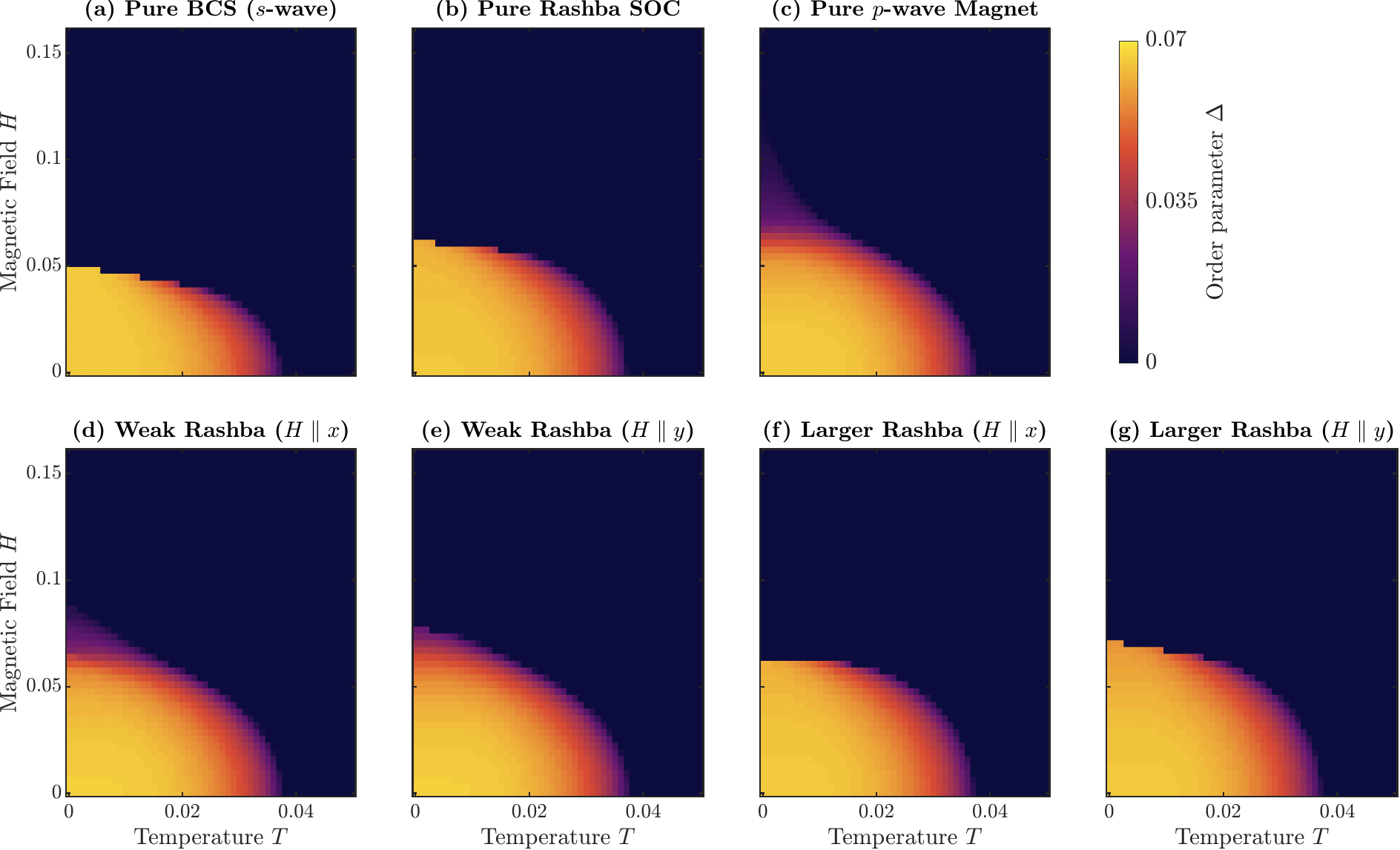}
\caption{Global $\Delta(T,H)$ phase diagrams. Top row: baseline configurations. Bottom row: hybridized configurations.}
\label{fig:global_phase}
\end{figure}

The order-parameter maps $\Delta(T,H)$ presented in Fig.~\ref{fig:global_phase} illustrate the thermodynamic evolution of the uniform superconducting state. At zero magnetic field, the zero-temperature gap $\Delta_0$ and critical temperature $T_c$ remain approximately unchanged across the configurations studied, consistent with the conventional BCS relation $\Delta_0 \approx 1.76 k_B T_c$. However, their responses to an applied field differ. Both the pristine $s$-wave [Fig.~\ref{fig:global_phase}(a)] and pure-Rashba [Fig.~\ref{fig:global_phase}(b)] systems undergo a first-order Pauli transition at low temperatures. For the parameters shown, Rashba SOC enhances the critical field while retaining the first-order character. The pure $p$-wave magnetic state [Fig.~\ref{fig:global_phase}(c)] instead exhibits a continuous second-order transition and a larger critical-field enhancement.

The coexistence of magnetic exchange and Rashba SOC produces a response that depends on the field orientation. In the weak-Rashba regime, the pairing amplitude vanishes continuously under $\mathbf{H}\parallel x$ [Fig.~\ref{fig:global_phase}(d)], whereas its suppression is sharper and the low-temperature critical field is lower under $\mathbf{H}\parallel y$ [Fig.~\ref{fig:global_phase}(e)]. In the strong-Rashba regime, an abrupt first-order collapse occurs under $\mathbf{H}\parallel x$ [Fig.~\ref{fig:global_phase}(f)]. The $\mathbf{H}\parallel y$ response also exhibits sharp order-parameter suppression near the low-temperature boundary [Fig.~\ref{fig:global_phase}(g)], but its critical field exceeds that for $\mathbf{H}\parallel x$.

These results show that the interplay between magnetic exchange and Rashba SOC modifies both the critical field and the depairing behavior differently for the two field orientations. This direction-dependent thermodynamic evolution accompanies the reversal from $H_{c2x}>H_{c2y}$ to $H_{c2y}>H_{c2x}$ discussed in Sec.~\ref{sec:phasediagram}. The maps do not include optimization over finite pair momentum.

\section{Asymptotic Behaviors in the Pure $p$-wave and Strong Rashba Limits}
\label{app:asymptotic}

\subsection{Nonlinear Enhancement of the Upper Critical Field in the Pure $p$-wave Limit}

In the main text, we show that the pure $p$-wave exchange provides out-of-plane spin-locking protection, producing an isotropic enhancement of the in-plane upper critical field. To characterize this baseline and its connection to the anisotropy reversal discussed in Sec.~\ref{sec:phasediagram}, we examine the dependence of $H_{c2}$ on the exchange strength $V_J$.

Figure~\ref{fig:asymptotic}(a) presents the normalized critical field $H_{c2}/H_{c20}$ as a function of $V_J$ in the pure $p$-wave limit ($V_R=0$), where $H_{c20}$ denotes the pristine $s$-wave reference critical field. The enhancement is nonlinear: for $V_J\lesssim0.04$, $H_{c2}$ increases only marginally and remains close to the pristine Pauli limit, indicating weak exchange-induced protection. At larger $V_J$, the critical field grows approximately linearly with the exchange strength. We refer to this parameter range as the ``linear protection regime,'' as indicated in Fig.~\ref{fig:asymptotic}(a). The crossover near $V_J\sim0.04$ thus marks the development of appreciable exchange-induced protection for the parameters studied. It is a sampled crossover, not an exact onset threshold or a region in which the exchange has identically zero effect.

This exchange dependence correlates with the parameter range supporting anisotropy reversal. For weak but finite exchange, $0<V_J\lesssim0.04$, no region with $H_{c2x}>H_{c2y}$ is resolved within the sampled finite-$V_R$ range. At larger $V_J$, an initial $H_{c2x}>H_{c2y}$ regime appears before the contrast reverses as Rashba SOC increases. This correspondence supports the interpretation that appreciable exchange-induced protection provides the baseline for the directional modification by weak Rashba SOC discussed in the main text.

\subsection{Asymptotic Behavior in the Strong Rashba Limit}

\begin{figure}[htbp]
\centering
\includegraphics[width=\linewidth]{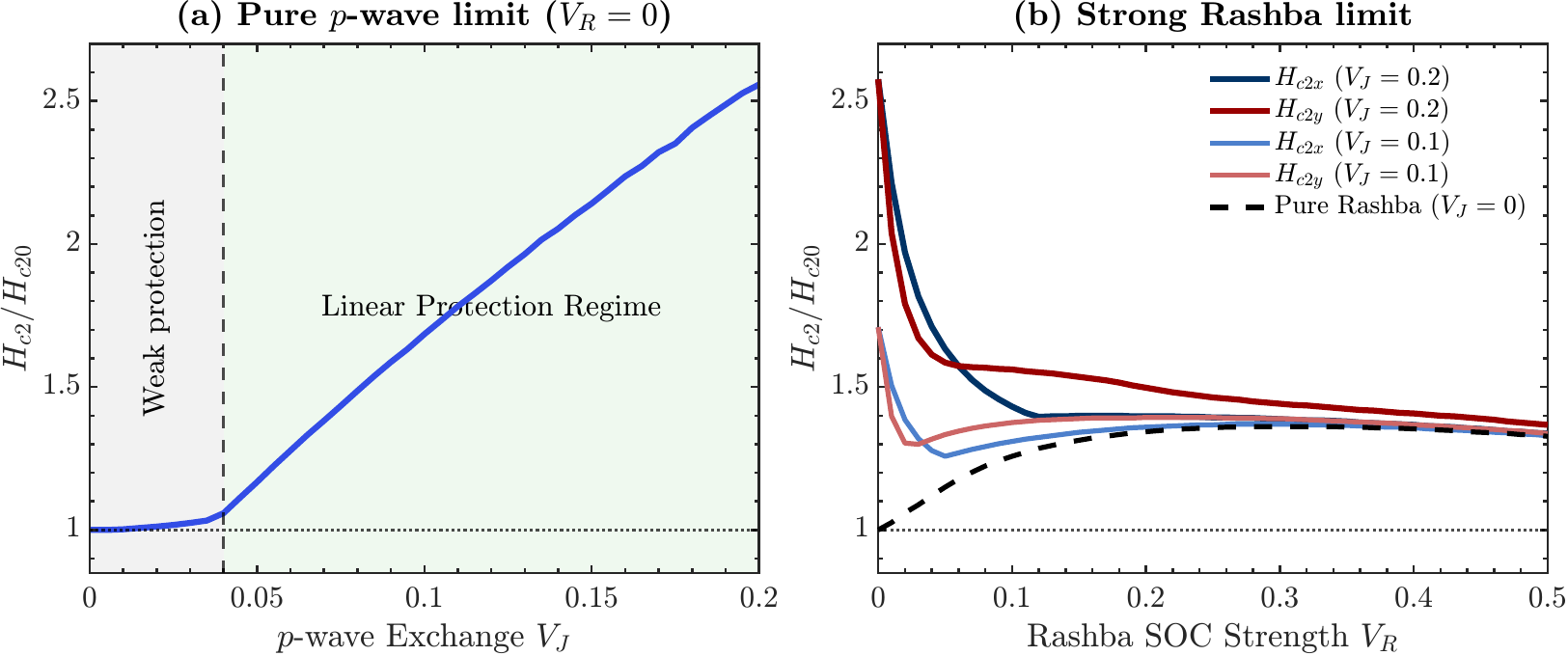}
\caption{Asymptotic behaviors of the normalized upper critical field $H_{c2}/H_{c20}$. \textbf{(a)} Nonlinear enhancement as a function of the $p$-wave exchange coupling $V_J$ in the pure $p$-wave limit ($V_R = 0$). The enhancement is strongly nonlinear: marginal for $V_J \lesssim 0.04$, crossing over into a quasi-linear regime at larger $V_J$. The weak-protection label denotes a crossover regime, not an exact threshold. \textbf{(b)} Evolution of $H_{c2x}$ and $H_{c2y}$ over an extended range of $V_R$. Solid curves ($V_J = 0.1, 0.2$) approach the pure-Rashba baseline (dashed) at large $V_R$. The normalizing field in both panels is the pristine $s$-wave reference $H_{c20}$, as in the main text.}
\label{fig:asymptotic}
\end{figure}

In the strong-Rashba limit ($V_R \gg V_J$), the twofold critical-field anisotropy induced by the $p$-wave exchange is progressively suppressed. As shown in Fig.~\ref{fig:asymptotic}(b), the principal-axis critical fields for $V_J = 0.1$ and $V_J = 0.2$ approach the pure-Rashba baseline ($V_J = 0$) as $V_R$ increases, with the differences substantially reduced for $V_R \gtrsim 0.4$. This behavior supports the interpretation that the anisotropy reversal occurs in an intermediate coupling regime where both magnetic exchange and Rashba SOC contribute appreciably to the superconducting field response.

\section{Microscopic Origin and Thermodynamic Control Model of the $p$-wave Form Factor}
\label{app:microscopic}

\begin{figure}[htbp]
\centering
\includegraphics[width=0.5\linewidth]{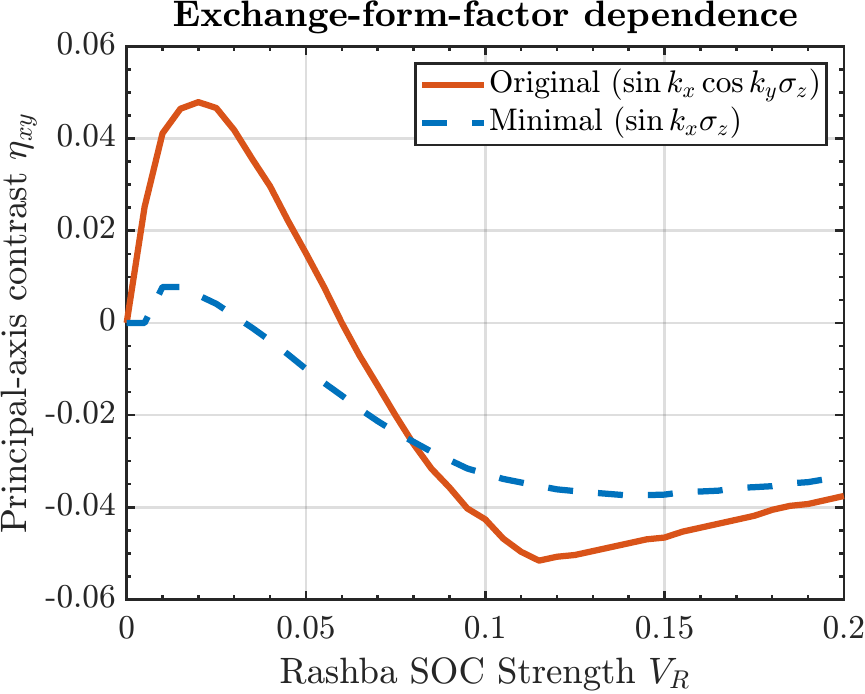}
\caption{Comparison of $\eta_{xy}$ for original form factor in the main text ($\sin k_x \cos k_y \sigma_z$, solid) and simplified form factor ($\sin k_x \sigma_z$, dashed).}
\label{fig:control}
\end{figure}

The effective $p$-wave magnetic Hamiltonian $h_J(\mathbf{k}) = V_J \gamma(\mathbf{k})\sigma_z$ originates from the coupling between an antiferromagnetic substrate featuring inter-sublattice hopping and a directional spin-orbit coupling (SOC)~\cite{Sun2025_PseudoIsing}. 
%The specific form factor $\gamma(\mathbf{k})$ can be explicitly derived by projecting a microscopic lattice model and applying perturbation theory.
We obtain its leading momentum dependence through second-order perturbation theory and adopt a simplified periodic lattice representation.
Specifically, the basic Hamiltonian is given by $H = H_{t} + H_{A} + H_{S}$, where the unperturbed kinetic energy $H_{t}$ and the perturbation terms are:
\begin{align}
    H_{t} &= 2t \left( \cos\frac{k_{x}}{2}\tau_{x} + \cos k_{y}\tau_{0} \right), \\
    H_{A} &= h_{A}\tau_{z} = \left( 2t_{A}\cos k_{y}\sigma_{y} \right) \tau_{z}, \\
    H_{S} &= h_{S}\tau_{y} = \left( 2t_{S}\sin\frac{k_{x}}{2}\sigma_{x} \right) \tau_{y}.
\end{align}
Treating the latter two terms as perturbations, we project the system into the upper ($\vert{}u\rangle$) and lower ($\vert{}d\rangle$) eigenstates of $H_t$. The corresponding band dispersion is $E_{u,d} = 2t\left( \cos k_{y} \pm \cos\frac{k_{x}}{2} \right)$, with an energy gap $\Delta E = 4t\cos\frac{k_{x}}{2}$.
Using second-order L\"owdin perturbation theory, the effective Hamiltonian for the low-energy upper band is obtained by integrating out the lower band. Evaluating the cross terms yields a commutator $i [h_{S}, h_{A}]$ that isolates the spin-splitting effective field:
\begin{equation}
    H_{\text{eff}} \approx E_{u} + \frac{h_{S}^{2} + h_{A}^{2} + i [h_{S}, h_{A}]}{\Delta E}.
\end{equation}
By neglecting the scalar quadratic terms, expanding the momenta near the relevant symmetry points, and incorporating the chemical potential $\mu$, we adopt the compact form $H_{\text{eff}} \approx h_{0}(\mathbf{k})\sigma_{0} + V_J \gamma(\mathbf{k})\sigma_{z}$. %Crucially, the emergent form factor $\gamma(\mathbf{k})$ is directly proportional to the cross product of the perturbation amplitudes, $h_{S} h_{A} \propto \sin\frac{k_x}{2}\cos k_y$. which maps to $\sin k_x \cos k_y$ in the continuum limit. 
Retaining the band-separation denominator, the projected exchange term is proportional to $\tan(k_x/2)\cos k_y$. Near $k_x=0$, this reduces to a term proportional to $k_x\cos k_y$, which we represent by the simplified periodic lattice form $\gamma(\mathbf{k})=\sin k_x\cos k_y$.
Here, the $\cos k_y$ dependence in $h_{A}$ captures the itinerant nature and bond-dependent exchange interactions of the underlying antiferromagnetic substrate~\cite{Sun2025_PseudoIsing}.
% If the antiferromagnetic background were instead to originate from a strictly localized, Ising-type on-site magnetization ($H_{A} = 2t_{A}\sigma_{y}\tau_{z}$), the effective spatial modulation would degenerate into a purely one-dimensional form factor $\gamma(\mathbf{k}) = \sin k_x$, entirely losing the $\cos k_y$ component associated with the transverse inter-site hopping.\\
 If the staggered exchange were instead purely on-site, $H_A=2t_A\sigma_y\tau_z$, the projected exchange would lose the $\cos k_y$ factor, and the corresponding lattice approximation would become $\gamma(\mathbf{k})=\sin k_x$.\\
To examine the macroscopic thermodynamic role of this 2D itinerant exchange form factor, we construct a control model in which the full modulation $\sin k_x \cos k_y \sigma_z$ is replaced by its localized-limit counterpart, $\sin k_x \sigma_z$. Both models retain the same two-dimensional lattice geometry.\\
As shown in Fig.~\ref{fig:control}, the control model also exhibits an anisotropy reversal from $\eta_{xy} > 0$ to $\eta_{xy} < 0$ as Rashba SOC increases. This demonstrates that the additional $\cos k_y$ modulation is not required for the reversal, supporting its interpretation in terms of the competition between odd-parity magnetic exchange and Rashba spin locking.\\
Nevertheless, the original form factor produces a substantially larger positive peak and a more pronounced negative minimum of $\eta_{xy}$ for the parameters shown. The $\cos k_y$ factor modulates the out-of-plane exchange strength along $k_y$ and introduces additional exchange nodes, changing the distribution of spin-locking protection across momentum space. This modification affects how the exchange texture competes with the Rashba field and can enhance the resulting directional contrast. The comparison therefore establishes that the exchange form factor strongly influences the magnitude of the critical-field anisotropy while preserving its reversal.

\section{Numerical Grid Independence and Temperature Evolution}
\label{app:convergence}

\begin{figure}[htbp]
\centering
\includegraphics[width=0.9\linewidth]{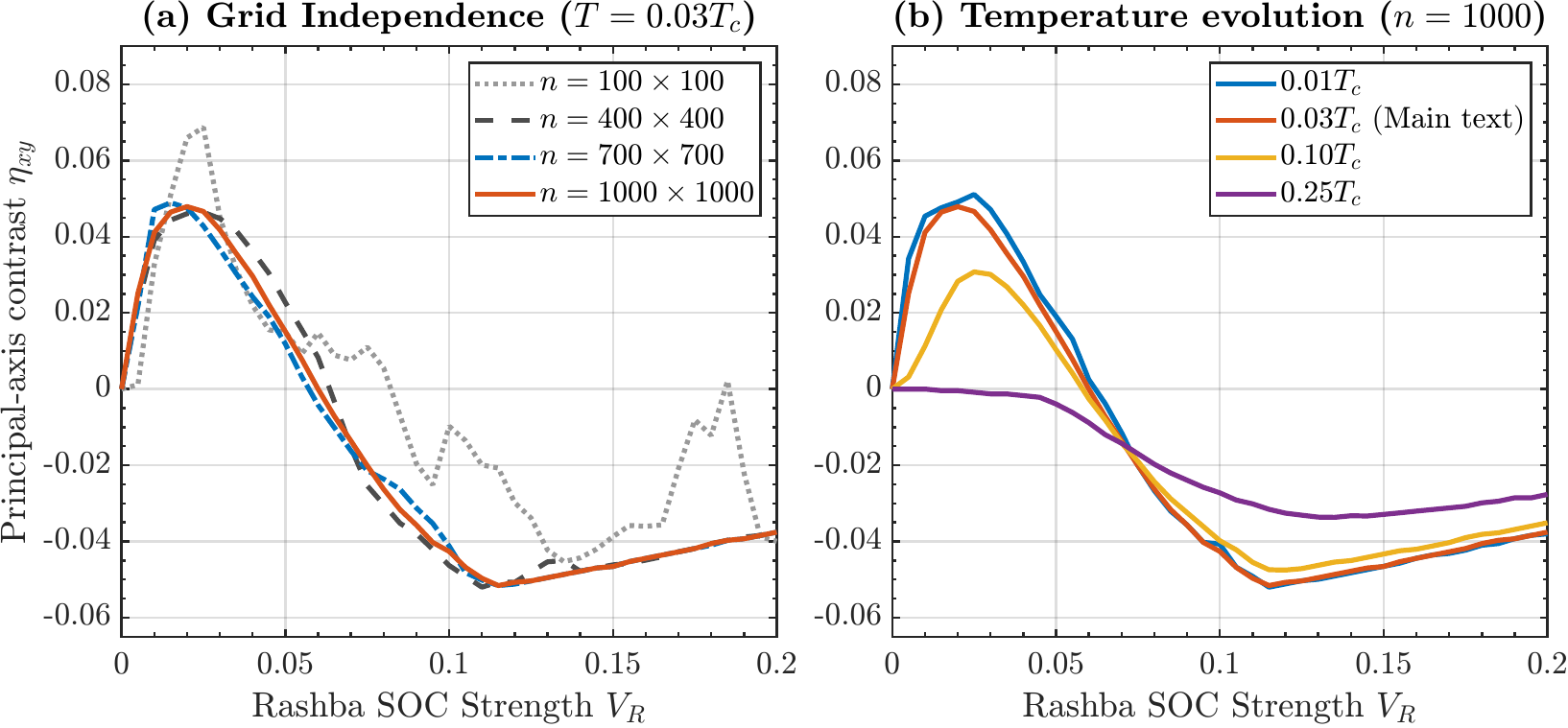}
\caption{(a) Grid independence test for $\eta_{xy}$ at $T = 0.03 T_c$. (b) Temperature evolution of $\eta_{xy}$ on the $1000 \times 1000$ grid.}
\label{fig:convergence}
\end{figure}

To assess convergence with respect to momentum-space sampling, we compared the principal-axis anisotropy ratio $\eta_{xy}$ on different $n\times n$ grids. As shown in Fig.~\ref{fig:convergence}(a), the results for $n\ge700$ exhibit similar angular-contrast evolution and consistently reproduce the sign reversal, supporting the robustness of the main result obtained using the $1000\times1000$ grid.

Figure~\ref{fig:convergence}(b) shows the temperature dependence of $\eta_{xy}$. The curves at $T=0.01T_c$ and $0.03T_c$ display similar anisotropy reversals. Increasing temperature generally weakens the anisotropy, and the positive-$\eta_{xy}$ region is no longer clearly resolved at $T=0.25T_c$. We therefore use $T=0.03T_c$ as a representative low temperature at which the reversal remains well resolved.

\newpage

\twocolumngrid % <--- 强制恢复双栏排版
\bibliographystyle{apsrev4-2}
\bibliography{refs}

\end{document}